\documentclass[reprint,onecolumn,notitlepage,showpacs,preprintnumbers,nofootinbib,amsmath,amssymb,aps,prd,floatfix,superscriptaddress]{revtex4-1}
\pdfoutput=1
\usepackage{graphicx}
\usepackage{bm}
\usepackage{subfigure}
\usepackage{url}
\usepackage[hyperindex]{hyperref}
\usepackage[svgnames]{xcolor}
\usepackage[ddmmyy,24hr]{datetime}
\usepackage{bigdelim}
\usepackage{booktabs}
\usepackage{dcolumn}
\usepackage{multirow}
\usepackage{subfigure}
\usepackage{cancel}
\usepackage{stackrel}
\usepackage{paralist}
\usepackage{xspace}
\usepackage{slashed}
\usepackage{ulem}
\newcommand{\nua}[1]{\ensuremath{\rlap{\kern-2.5pt\ensuremath{\overset{\scriptscriptstyle(-)}{\phantom{\nu}}}}{\ensuremath{{\nu}_{#1}}}}}

\usepackage{enumitem}

\begin{document}

\title{Neutrino-4 anomaly: oscillations or fluctuations?}

\author{C. Giunti}
\email{carlo.giunti@to.infn.it}
\affiliation{Istituto Nazionale di Fisica Nucleare (INFN), Sezione di Torino, Via P. Giuria 1, I--10125 Torino, Italy}

\author{Y.F. Li}
\email{liyufeng@ihep.ac.cn}
\affiliation{Institute of High Energy Physics,
Chinese Academy of Sciences, Beijing 100049, China}
\affiliation{School of Physical Sciences, University of Chinese Academy of Sciences, Beijing 100049, China}

\author{C.A. Ternes}
\email{ternes@to.infn.it}
\affiliation{Istituto Nazionale di Fisica Nucleare (INFN), Sezione di Torino, Via P. Giuria 1, I--10125 Torino, Italy}

\author{Y.Y. Zhang}
\email{zhangyiyu@ihep.ac.cn}
\affiliation{Institute of High Energy Physics,
Chinese Academy of Sciences, Beijing 100049, China}
\affiliation{School of Physical Sciences, University of Chinese Academy of Sciences, Beijing 100049, China}

%\date{\dayofweekname{\day}{\month}{\year} \ddmmyydate\today, \currenttime}
\date{4 March 2021}

\begin{abstract}
We present a deep study of the Neutrino-4 data
aimed at finding the statistical significance of the
large-mixing
short-baseline neutrino oscillation signal claimed by the Neutrino-4
collaboration at more than $3\sigma$.
We found that the results of the Neutrino-4
collaboration can be reproduced approximately only by neglecting the
effects of the energy resolution of the detector.
Including these effects,
we found that the best fit is obtained for a mixing that is even larger,
close to maximal,
but the statistical significance of the
short-baseline neutrino oscillation signal is only about
$2.7\sigma$
if evaluated with the usual method based on Wilks' theorem.
We show that the large Neutrino-4 mixing is in strong tension
with the
KATRIN,
PROSPECT,
STEREO, and
solar $\nu_{e}$
bounds.
Using a more reliable Monte Carlo simulation of a large set of Neutrino-4-like data,
we found that the
statistical significance of the
Neutrino-4 short-baseline neutrino oscillation signal decreases to about
$2.2\sigma$.
We also show that it is not unlikely to find
a best-fit point that has a large mixing,
even maximal,
in the absence of oscillations.
Therefore,
we conclude that the claimed Neutrino-4 indication in favor of short-baseline neutrino oscillations with very large mixing is rather doubtful.
\end{abstract}

%\pacs{}

\maketitle

\section{Introduction}
\label{sec:Introduction}

The possible existence of sterile neutrinos is
one of the current hot research topics in the quest for new physics beyond the Standard Model.
In particular,
the possible existence of light sterile neutrinos at the eV mass scale
has been motivated by the
LSND~\cite{Aguilar:2001ty},
MiniBooNE~\cite{Aguilar-Arevalo:2018gpe,Aguilar-Arevalo:2020nvw},
Gallium~\cite{Abdurashitov:2005tb,Laveder:2007zz,Giunti:2006bj},
and
reactor~\cite{Mention:2011rk}
anomalies,
that may be a signal of short baseline active-sterile neutrino oscillations
(see the recent reviews in Refs.~\cite{Giunti:2019aiy,Diaz:2019fwt,Boser:2019rta}).
Recently, a strong indication in favor of such oscillations
has been claimed by the Neutrino-4
collaboration~\cite{Serebrov:2018vdw,Serebrov:2020rhy,Serebrov:2020kmd}.
However, this result is controversial and has been criticized in
Ref.~\cite{Almazan:2020drb}
(see also the answer in Ref.~\cite{Serebrov:2020yvp})
and in Ref.~\cite{Danilov-Skrobova-2020-JETPL-112-452}.

In this paper we present a detailed analysis of the Neutrino-4 data
and we discuss in particular the effect of the energy resolution of the detector,
that seems not to have been taken into account
in the analysis of the Neutrino-4 collaboration
(as independently noted in Refs.~\cite{Danilov:2018dme,Danilov-Skrobova-2020-JETPL-112-452}).
We also present the results of a Monte Carlo implementation of the
statistical analysis of the Neutrino-4 data that is more reliable than the standard
methods based on the $\chi^2$ distribution
predicted by Wilks' theorem~\cite{Wilks:1938dza},
as discussed in
Refs.~\cite{Feldman:1997qc,Lyons:2014kta,Agostini:2019jup,Algeri:2019arh,Giunti:2020uhv,Coloma:2020ajw}.

In Section~\ref{sec:method}
we present our analysis method,
in Section~\ref{sec:Wilks}
we discuss the results that we obtained using the standard statistical method
based on Wilks' theorem,
in Section~\ref{sec:MC}
we present the results obtained with a Monte Carlo statistical analysis,
and finally in Section~\ref{sec:Conclusions}
we summarize our results and draw our conclusions.

\section{Analysis method}
\label{sec:method}

The Neutrino-4 experiment is operating since 2014 close to the SM-3 reactor
in Dimitrovgrad (Russia).
It is a 90 MW research reactor with a compact core
($42\times42\times35 \, \text{cm}^3$)
using highly enriched $^{235}\text{U}$ fuel.
The Neutrino-4 detector is made of liquid scintillator
with 0.1\% gadolinium concentration for a better detection of the neutron in the
$ \bar\nu_{e} + p \to n + e^{+} $
detection process of the reactor $\bar\nu_{e}$'s.
The detector is segmented in 50 sections with active size
$22.5\times22.5\times70 \, \text{cm}^3$
disposed in 10 rows.
The detector can move horizontally, on the side of the reactor,
allowing measurements of the $\bar\nu_{e}$ at distances between
6.4 m and 11.9 m from the center of the reactor.
The detector recorded about 300 $\bar\nu_{e}$ events per day
at the average distance of 8 m from the reactor core.
The last published data~\cite{Serebrov:2020rhy,Serebrov:2020kmd}
correspond to 720 days of reactor-on data taking
and
417 days of reactor-off background measurements.

The data have been taken by the Neutrino-4 collaboration at
$ n_{L} = 24 $ distances between
6.4 m and 11.9 m from the center of the reactor,
at intervals of 23.5 cm,
and divided in
$ n_{E} = 9 $ bins of prompt energy from 1.5 to 6 MeV
with uniform $\Delta{E} = 500 \, \text{keV} $ width.
Let us remind that the neutrino energy
$E$
is related to the prompt energy
$E_{\text{p}}$
by
\begin{equation}
E
=
E_{\text{p}} + m_{n} - m_{p} - m_{e}
\simeq
E_{\text{p}} + 0.78 \, \text{MeV}
.
\label{Ep}
\end{equation}
The 500 keV prompt energy bin width was motivated by the calibration of the detector
with a $^{22}\text{Na}$ $\gamma$-source
that showed an energy resolution of
about 142 keV at the 511 keV $^{22}\text{Na}$ line
and
about 217 keV at the 1274 keV $^{22}\text{Na}$ line
(see Figure~22 of Ref.~\cite{Serebrov:2020kmd}).
It was also calibrated with the process
$ n + p \to d + \gamma $
that showed an energy resolution of
about 276 keV at 2.2 MeV
(see Figure~21 of Ref.~\cite{Serebrov:2020kmd}).

The Neutrino-4 collaboration collected the data in the 216 ratios
\begin{equation}
R_{ik}^{\text{exp}}
=
\dfrac
{ N_{ik}^{\text{exp}} L_{k}^2 }
{ n_{L}^{-1} \sum_{k'=1}^{n_{L}} N_{ik'} L_{k'}^2 }
,
\label{Rik-exp}
\end{equation}
where $i=1,\ldots,n_{E}$ is the neutrino energy index,
$k=1,\ldots,n_{L}$ is the reactor-detector distance index,
$L_{k}$ is the average distance of the $k^{\text{th}}$ distance interval,
and $N_{ik}^{\text{exp}}$ is the rate of observed $\bar\nu_{e}$ events
in the $i^{\text{th}}$ energy bin and $k^{\text{th}}$ distance interval.

The results of short-baseline neutrino oscillation experiments
as Neutrino-4
are typically analyzed in the framework of 3+1 active-sterile neutrino mixing.
This is the simplest extension of standard three-neutrino mixing
that can explain short-baseline neutrino oscillations
with minimal perturbations to the standard three-neutrino mixing global fit of
solar, atmospheric and long-baseline (accelerator and reactor) neutrino oscillation
data~\cite{Capozzi:2017ipn,Esteban:2020cvm,deSalas:2020pgw}.
In the 3+1 framework it is assumed that
there is a non-standard massive neutrino $\nu_{4}$
with mass $m_{4} \gtrsim 1 \, \text{eV}$
such that
$ m_{1}, m_{2}, m_{3} \ll m_{4} $,
where $ m_{1}, m_{2}, m_{3} $ are the masses of the three standard massive neutrinos
$ \nu_{1}, \nu_{2}, \nu_{3} $.
In the flavor basis,
besides the three standard active neutrinos
$ \nu_{e}, \nu_{\mu}, \nu_{\tau} $
there is a sterile neutrino $\nu_{s}$,
and the mixing is given by
\begin{equation}
\nu_{\alpha}
=
\sum_{j=1}^{4} U_{\alpha j} \nu_{j}
\qquad
(\alpha=e,\mu,\tau,s)
,
\label{mixing}
\end{equation}
where $U$ is the $4\times4$ unitary mixing matrix.
In this 3+1 framework,
the effective short-baseline (SBL) survival probability
of electron neutrinos and antineutrinos is given by
\begin{equation}
P_{ee}^{\text{SBL}}
=
1 - \sin^2\!2\vartheta_{ee} \, \sin^2\!\left( \dfrac{\Delta{m}^2_{41} L}{4 E} \right)
,
\label{Pee}
\end{equation}
where
$\Delta{m}^2_{41}=m_{4}^2-m_{1}^2$
and
$\sin^2\!2\vartheta_{ee} = 4 |U_{e4}|^2 ( 1 - |U_{e4}|^2 )$,
where $\vartheta_{ee}$ is an effective mixing angle
that coincides with $\vartheta_{14}$
in the standard parameterization of the
mixing matrix
(see the recent reviews in Refs.~\cite{Giunti:2019aiy,Diaz:2019fwt,Boser:2019rta}).

In the framework of 3+1 neutrino mixing,
the expected rate of events
in the $i^{\text{th}}$ energy bin and $k^{\text{th}}$ distance interval
of the Neutrino-4 experiment is given by
\begin{equation}
N_{ik}^{\text{the}}
=
\dfrac{N_{i}^{0}}{L_{k}^2}
\left[
1
-
\sin^2\!2\vartheta_{ee}
\left\langle \sin^2\left(\frac{\Delta{m}^2_{41}L}{4E}\right) \right\rangle_{ik}
\right]
,
\label{Nik-the}
\end{equation}
where $N_{i}^{0}$ is the rate of expected events
in the $i^{\text{th}}$ energy bin without
short-baseline neutrino oscillations
at a reference distance where $L_{k}=1$,
and the oscillating terms are averaged
over the appropriate neutrino energy range and the distance uncertainty
of the $i^{\text{th}}$ energy bin and $k^{\text{th}}$ distance interval.
Therefore,
the theoretical prediction $R_{ik}^{\text{the}}$
for the ratios in Eq.~(\ref{Rik-exp}) is given by
\begin{equation}
R_{ik}^{\text{the}}
=
\dfrac{
1
-
\sin^2\!2\vartheta_{ee}
\left\langle \sin^2\left(\frac{\Delta{m}^2_{41}L}{4E}\right) \right\rangle_{ik}
}{
1
-
\sin^2\!2\vartheta_{ee}
\,
n_{L}^{-1} \sum_{k'=1}^{n_{L}}
\left\langle \sin^2\left(\frac{\Delta{m}^2_{41}L}{4E}\right) \right\rangle_{ik'}
}
.
\label{Rik-the}
\end{equation}
This expression shows the advantage of considering the ratios in Eq.~(\ref{Rik-exp}),
which must be confronted with the theoretical ratios
$R_{ik}^{\text{the}}$ that do not depend on the values of rates $N_{i}^{0}$,
whose estimation has large uncertainties,
mainly due to the uncertainties of the
theoretical reactor neutrino fluxes
(see the reviews in Refs.~\cite{Huber:2016fkt,Hayes:2016qnu}).

\begin{figure*}[!t]
\centering
\includegraphics*[width=\linewidth]{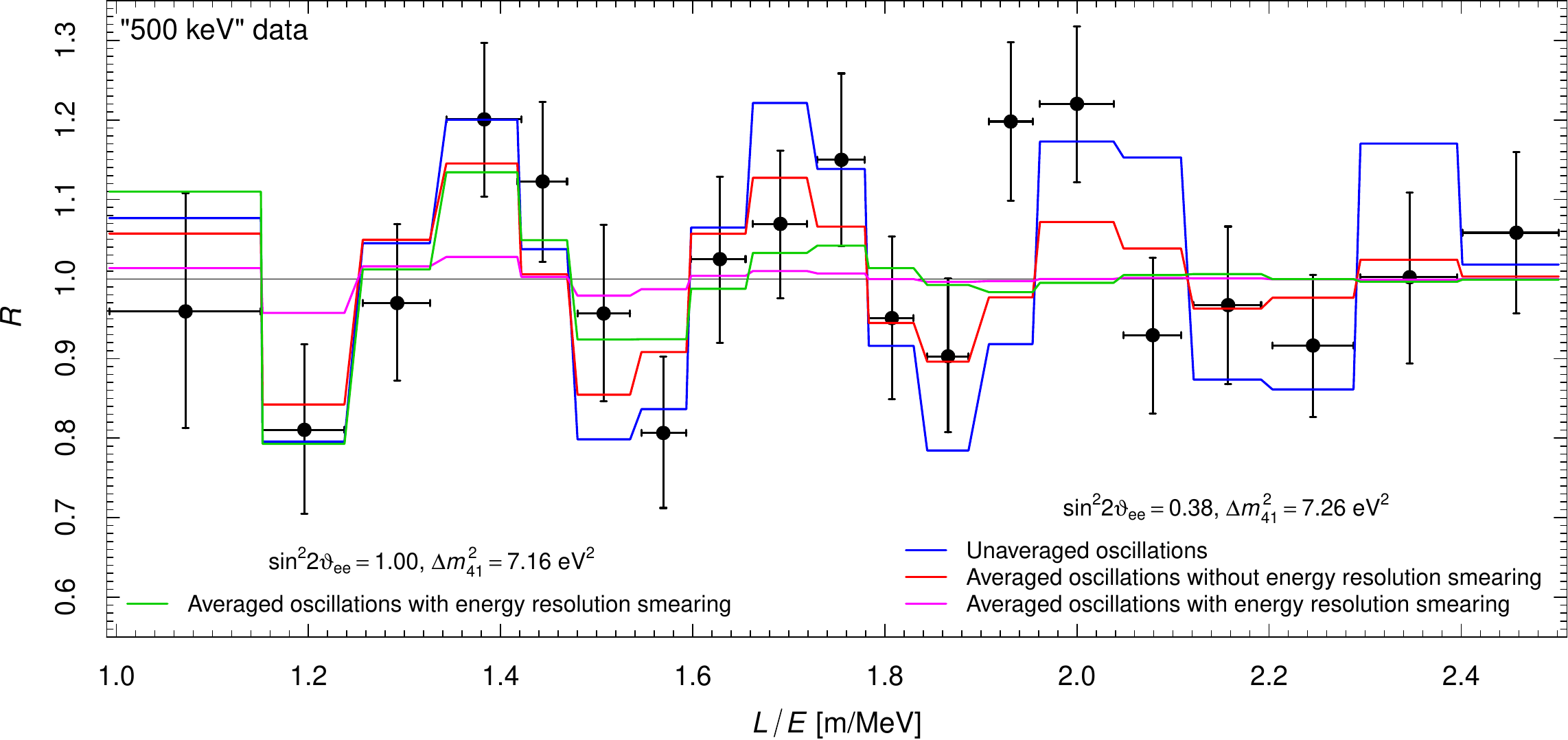}
\caption{ \label{fig:500keV}
Values of the
"500 keV" experimental ratios $R_{j}^{\text{exp}}$
(black points with error bars).
The blue, red, and magenta histograms
show, respectively,
the values of the theoretical ratios
$R_{j}^{\text{the}}$
obtained
for
$ \sin^2\!2\vartheta_{ee} = 0.38 $
and
$ \Delta{m}^2_{41} = 7.26 \, \text{eV}^2 $
without any averaging of the oscillating terms,
with averaging over the energy and distance intervals without energy resolution,
and
with averaging over the energy and distance intervals with energy resolution.
The green histogram
corresponds to the best-fit values
$ \sin^2\!2\vartheta_{ee} = 1 $
and
$ \Delta{m}^2_{41} = 7.16 \, \text{eV}^2 $
obtained
with averaging over the energy and distance intervals with energy resolution.
}
\end{figure*}

The Neutrino-4 collaboration further processed the data
by dividing the $ n_{\text{tot}} = 216 $ $R_{ik}^{\text{exp}}$'s
in groups of $ n_{g} = 8 $ values that correspond to neighboring $L/E$ intervals,
arguing that the $L/E$ dependence of the ratios $R_{ik}^{\text{exp}}$
``allows the direct demonstration of the effect of oscillations''.
In this way,
they obtained the $ n_{L/E} = 27 $ averages
\begin{equation}
R_{j}^{\text{exp}}
=
\dfrac{1}{n_{g}}
\sum_{ i,k \in g(j)}
R_{ik}^{\text{exp}}
=
\dfrac{1}{n_{g}}
\sum_{ i,k \in g(j)}
\dfrac
{ N_{ik}^{\text{exp}} L_{k}^2 }
{ n_{L}^{-1} \sum_{k'=1}^{n_{L}} N_{ik'} L_{k'}^2 }
,
\label{Rjexp}
\end{equation}
where $g(j)$ is the set of indices $i$ and $k$ that corresponds to the $j^{\text{th}}$
group of neighboring $L/E$ intervals.
The Neutrino-4 collaboration published only the first $ n'_{L/E} = 19 $
values of the ratios $R_{j}^{\text{exp}}$
in Figure~52 of Ref.~\cite{Serebrov:2020kmd}
("500 keV" data),
corresponding to a range of $L/E$ from about 1 to about 2.5 m/MeV,
arguing that for $ L/E > 2.5 \text{m/MeV} $ the oscillations
are averaged-out by the energy resolution of the detector.
We reproduced these data and their uncertainties
in Figure~\ref{fig:500keV}.

The Neutrino-4 collaboration presented also the results of an average of
the $R_{j}^{\text{exp}}$'s obtained with the
500 keV energy bin width discussed above
and similar $R_{j}^{\text{exp}}$'s
obtained with the
125 and 250 keV energy bin widths,
arguing that the average of different sampling of the data
suppresses fluctuations.
We think that this procedure is rather ad-hoc and inappropriate,
since the energy resolution of the detector is larger than about 200 keV.
This procedure was criticized also in Ref.~\cite{Danilov-Skrobova-2020-JETPL-112-452}
on the basis of similar arguments.
Therefore,
in this paper we consider first the Neutrino-4 "500 keV" data
and then we discuss how the results change by considering
the "125-250-500 keV" averaged data.

The averages $R_{j}^{\text{exp}}$ must be compared with the
corresponding theoretical averages
\begin{equation}
R_{j}^{\text{the}}
=
\dfrac{1}{n_{g}}
\sum_{ i,k \in g(j)}
R_{ik}^{\text{the}}
=
\dfrac{1}{n_{g}}
\sum_{ i,k \in g(j)}
\dfrac{
1
-
\sin^2\!2\vartheta_{ee}
\left\langle \sin^2\left(\frac{\Delta{m}^2_{41}L}{4E}\right) \right\rangle_{ik}
}{
1
-
\sin^2\!2\vartheta_{ee}
\,
n_{L}^{-1} \sum_{k'=1}^{n_{L}}
\left\langle \sin^2\left(\frac{\Delta{m}^2_{41}L}{4E}\right) \right\rangle_{ik'}
}
.
\label{Rjthe}
\end{equation}
We calculated the averaged oscillation terms with
\begin{equation}
\left\langle \sin^2\left(\frac{\Delta{m}^2_{41}L}{4E}\right) \right\rangle_{ik}
=
\dfrac{
\int_{L_{k}^{\text{min}}}^{L_{k}^{\text{max}}} d L
\,
L^{-2}
\int_{E_{i}^{\text{min}}}^{E_{i}^{\text{max}}} d E'_{\text{p}}
\int d E_{\text{p}}
\,
R(E_{\text{p}},E'_{\text{p}})
\,
\sin^2\left(\frac{\Delta{m}^2_{41}L}{4E}\right)
\,
\phi_{\bar\nu_{e}}(E)
\,
\sigma_{ \bar\nu_{e} p }(E)
}
{
\int_{L_{k}^{\text{min}}}^{L_{k}^{\text{max}}} d L
\,
L^{-2}
\int_{E_{i}^{\text{min}}}^{E_{i}^{\text{max}}} d E'_{\text{p}}
\int d E_{\text{p}}
\,
R(E_{\text{p}},E'_{\text{p}})
\,
\phi_{\bar\nu_{e}}(E)
\,
\sigma_{ \bar\nu_{e} p }(E)
}
,
\label{ave}
\end{equation}
where
$\phi_{\bar\nu_{e}}(E)$ is the reactor neutrino flux,
$\sigma_{ \bar\nu_{e} p }(E)$ is the detection cross section,
$E$ and $E_{\text{p}}$ are related by Eq.~(\ref{Ep}),
$E_{i}^{\text{min}}$ and $E_{i}^{\text{max}}$
delimit the $i^{\text{th}}$ energy bin,
$L_{k}^{\text{min}}$ and $L_{k}^{\text{max}}$
delimit the $k^{\text{th}}$ distance interval,
and
$R(E_{\text{p}},E'_{\text{p}})$
is the energy resolution function
\begin{equation}
R(E_{\text{p}},E'_{\text{p}})
=
\dfrac{1}{\sqrt{2\pi}\sigma_{E_{p}}}
\,
\exp\left( - \dfrac{(E_{\text{p}}-E'_{\text{p}})^2}{2\sigma_{E_{p}}^2} \right)
,
\label{Res}
\end{equation}
with energy resolution $\sigma_{E_{p}}$ given by
\begin{equation}
\sigma_{E_{p}}
=
0.19 \, \sqrt{ \dfrac{E_{p}}{\text{MeV}} } \, \text{MeV}
.
\label{sigEp}
\end{equation}
The well-known dependence of the energy resolution on the square root of the energy
is due to the Poisson statistics of photoelectrons.
The coefficient is adapted to match the above-mentioned
calibrations of the detector shown in Figures~21 and~22 of Ref.~\cite{Serebrov:2020kmd}
and agrees with the independent estimation in
Ref.~\cite{Danilov-Skrobova-2020-JETPL-112-452}.
Note that other effects beyond the photon statistics, such as the detector leakage and non-uniformity, may contribute to the resolution function.
These effects are usually taken into account through an appropriate smearing matrix that is estimated by the experimental collaboration.
Since the Neutrino-4 collaboration did not provide such information,
we consider our analysis with the energy resolution function in Eq.~(\ref{Res})
the best working assumption that we can make
with the limited available information.
\begin{table*}[t!]
\centering
\renewcommand{\arraystretch}{1.2}
\begin{tabular}{ccccc}
Neutrino-4
&
\multicolumn{2}{c}{"500 keV" data}
&
\multicolumn{2}{c}{"125-250-500 keV" data}
\\[-0.3cm]
&
\multicolumn{2}{c}{\rule{2.5cm}{0.5pt}}
&
\multicolumn{2}{c}{\rule{2.5cm}{0.5pt}}
\\
&
{\renewcommand{\arraystretch}{0.8}
\begin{tabular}{c}
without
\\
en. res.
\end{tabular}
}
&
{\renewcommand{\arraystretch}{0.8}
\begin{tabular}{c}
with
\\
en. res.
\end{tabular}
}
&
{\renewcommand{\arraystretch}{0.8}
\begin{tabular}{c}
without
\\
en. res.
\end{tabular}
}
&
{\renewcommand{\arraystretch}{0.8}
\begin{tabular}{c}
with
\\
en. res.
\end{tabular}
}
\\
\hline
$\chi^{2}_{\text{min}}$
&
$14.9$
&
$18.2$
&
$21.9$
&
$21.1$
\\
GoF
&
$60\%$
&
$37\%$
&
$19\%$
&
$22\%$
\\
$(\sin^2\!2\vartheta_{ee})_{\text{bf}}$
&
$0.38$
&
$1.0$
&
$0.27$
&
$0.93$
\\
$(\Delta{m}^2_{41})_{\text{bf}}$
&
$7.2$
&
$7.2$
&
$8.8$
&
$7.2$
\\
$\Delta\chi^{2}_{\text{NO}}$
&
$13.1$
&
$9.8$
&
$9.9$
&
$10.7$
\\
\multicolumn{5}{c}{$\chi^2$ distribution}
\\
\hline
%$\Delta(\sin^2\!2\vartheta_{ee})_{3\sigma}$
%&
%$ 0.08 - 0.73 $
%&
%$ > 0.04 $
%&
%$ < 0.46 $
%&
%$ > 0.02 $
%\\
%$\Delta(\Delta{m}^2_{41})_{3\sigma}$
%&
%$ 5.11 - 8.98 $
%&
%$ 1.16 - 9.37 $
%&
%$ < 9.12 $
%&
%$ < 9.37 $
%\\
$p$-value
&
$0.0014$
&
$0.0075$
&
$0.0072$
&
$0.0048$
\\
$\sigma$-value
&
$3.2$
&
$2.7$
&
$2.7$
&
$2.8$
\\
\multicolumn{5}{c}{Monte Carlo distribution}
\\
\hline
%$\Delta(\sin^2\!2\vartheta_{ee})_{3\sigma}$
%&
%$ < 0.76 $
%&
%$-$
%&
%$ < 0.49 $
%&
%$-$
%\\
%$\Delta(\Delta{m}^2_{41})_{3\sigma}$
%&
%$-$
%&
%$-$
%&
%$-$
%&
%$-$
%\\
$p$-value
&
$0.011$
&
$0.028$
&
$0.087$
&
$0.026$
\\
$\sigma$-value
&
$2.5$
&
$2.2$
&
$1.7$
&
$2.2$
\end{tabular}
\caption{ \label{tab:fits}
Results of our fits of the Neutrino-4 "500 keV" and "125-250-500 keV" data:
minimum $\chi^2$ ($\chi^{2}_{\text{min}}$),
goodness of fit (GoF) for 17 degrees of freedom,
best fit values of
$\sin^2\!2\vartheta_{ee}$ and $\Delta{m}^2_{41}$, and
difference $\Delta\chi^{2}_{\text{NO}}$ between the $\chi^{2}$ of no oscillations and $\chi^{2}_{\text{min}}$.
The last four rows give the
$p$-value
and
the number of $\sigma$'s ($\sigma$-value)
corresponding to $\Delta\chi^{2}_{\text{NO}}$
obtained considering the $\chi^2$ distribution
(with two degrees of freedom
corresponding to two fitted oscillation parameters
$\sin^22\vartheta_{ee}$ and $\Delta{m}^2_{41}$)
and with a Monte Carlo estimation of the true distribution.
}
\end{table*}

\section{Wilks' confidence intervals}
\label{sec:Wilks}

In this Section we present the results that we obtained from the fits of the
"500 keV" 
and
"125-250-500 keV" data
considering the standard $\chi^2$ distribution
for log-likelihood ratios,
as predicted by Wilks' theorem~\cite{Wilks:1938dza}.
We perform the fit of the Neutrino-4 data with the least-squares function
\begin{equation}
\chi^2
=
\sum_{j=1}^{19}
\left(
\dfrac{ R_{j}^{\text{the}} - R_{j}^{\text{exp}} }{ \Delta{R}_{j}^{\text{exp}} }
\right)^2
,
\label{chi2}
\end{equation}
where
$R_{j}^{\text{exp}}$
and
$\Delta{R}_{j}^{\text{exp}}$ are the
experimental data and their uncertainties.
The "500 keV" data, that we consider first,
are given in Figure~52 of Ref.~\cite{Serebrov:2020kmd}
and reproduced in Figure~\ref{fig:500keV}.

The Neutrino-4 collaboration fitted the "500 keV" data with the
least-squares function
\begin{equation}
\widetilde\chi^2
=
\sum_{i,k}
\left(
\dfrac{ R_{ik}^{\text{the}} - R_{ik}^{\text{exp}} }{ \Delta{R}_{ik}^{\text{exp}} }
\right)^2
,
\label{chi2tilde}
\end{equation}
and obtained a $3.5\sigma$ evidence of neutrino oscillations
with best-fit values
$ \sin^2\!2\vartheta_{ee} = 0.38 $
and
$ \Delta{m}^2_{41} = 7.26 \, \text{eV}^2 $.
The allowed regions in the
($\sin^2\!2\vartheta_{ee},\Delta{m}^2_{41}$)
plane obtained by the Neutrino-4 collaboration
from the "500 keV" data
are shown in Figure~50 of Ref.~\cite{Serebrov:2020kmd}.
One can see that
there are five
narrow-$\Delta{m}^2_{41}$ allowed regions
at $3\sigma$.

We cannot reproduce the Neutrino-4 analysis of the "500 keV" data using
the least-squares function (\ref{chi2tilde}),
because the values of the $R_{ik}^{\text{exp}}$ have not been published by the
Neutrino-4 collaboration.
However,
we can reproduce approximately the Neutrino-4 analysis of the "500 keV" data using
the least-squares function (\ref{chi2}).
In particular,
it is clear that the best-fit values of the oscillation parameters
found by the Neutrino-4 collaboration should fit well the experimental
$R_{j}^{\text{exp}}$
values.
Unfortunately,
this does not happen,
as one can see from Figure~\ref{fig:500keV},
where the corresponding theoretical $R_{j}^{\text{the}}$ values are represented
by the magenta histogram.
The reason is that the energy resolution suppresses strongly the oscillations,
especially at small values of $L/E$,
where $\sigma_{E_{p}}$ is larger because the energy is large.
For example,
for the maximal prompt energy of 6 MeV,
that corresponds to values of $L/E$ in the first bin at
$ L/E \simeq 1.1 \, \text{m/MeV} $,
from Eq.~(\ref{sigEp}) the energy resolution is about 
0.46 MeV and the width of the Gaussian resolution function (\ref{Res})
is almost twice of the 500 keV bin width.

As shown in the second column in Table~\ref{tab:fits},
the best fit of the "500 keV" data taking into account the
energy resolution of the detector corresponds to maximal mixing
($\sin^2\!2\vartheta_{ee}=1$).
This value is much larger than the
best fit value $ \sin^2\!2\vartheta_{ee} = 0.38 $
found by the Neutrino-4 collaboration.
The corresponding theoretical $R_{j}^{\text{the}}$ values are shown by the green histogram
in Figure~\ref{fig:500keV}.

We can fit the "500 keV" data with the best-fit values of the oscillation parameters
found by the Neutrino-4 collaboration only by neglecting the energy resolution of the detector,
i.e. by considering
$ R(E_{\text{p}},E'_{\text{p}}) = \delta(E_{\text{p}}-E'_{\text{p}}) $.
Under this assumption,
we obtained the theoretical $R_{j}^{\text{the}}$ values represented
by the red histogram in Figure~\ref{fig:500keV},
that gives an acceptable fit of the data.
For comparison,
in Figure~\ref{fig:500keV} we have also drawn the blue
histogram corresponding to unaveraged oscillations,
that is obtained by replacing
$
\left\langle \sin^2\left(\frac{\Delta{m}^2_{41}L}{4E}\right) \right\rangle_{ik}
$
with
$
\sin^2\left(\frac{\Delta{m}^2_{41}L_{k}}{4E_{i}}\right)
$.
In this case,
in the calculation of
$R_{j}^{\text{the}}$
there is only the averaging due to the mean of the eight $R_{ij}^{\text{the}}$
with different $L_{k}$ and $E_{i}$
that contribute to each bin of $L/E$.

From the red histogram in Figure~\ref{fig:500keV}
one can see that the oscillations averaged over the
sizes of the energy bins and distance intervals
without the energy resolution smearing
are strongly suppressed at large values of $L/E$
with respect to unaveraged oscillations.
On the other hand,
for small values of $L/E$ the suppression is rather small
and one can fit the data with the best-fit values of the oscillation parameters
claimed by the Neutrino-4 collaboration.

\begin{figure*}[!t]
\centering
\setlength{\tabcolsep}{0pt}
\begin{tabular}{cc}
\subfigure[]{\label{fig:500-Wilks}
\begin{tabular}{c}
\includegraphics*[width=0.49\linewidth]{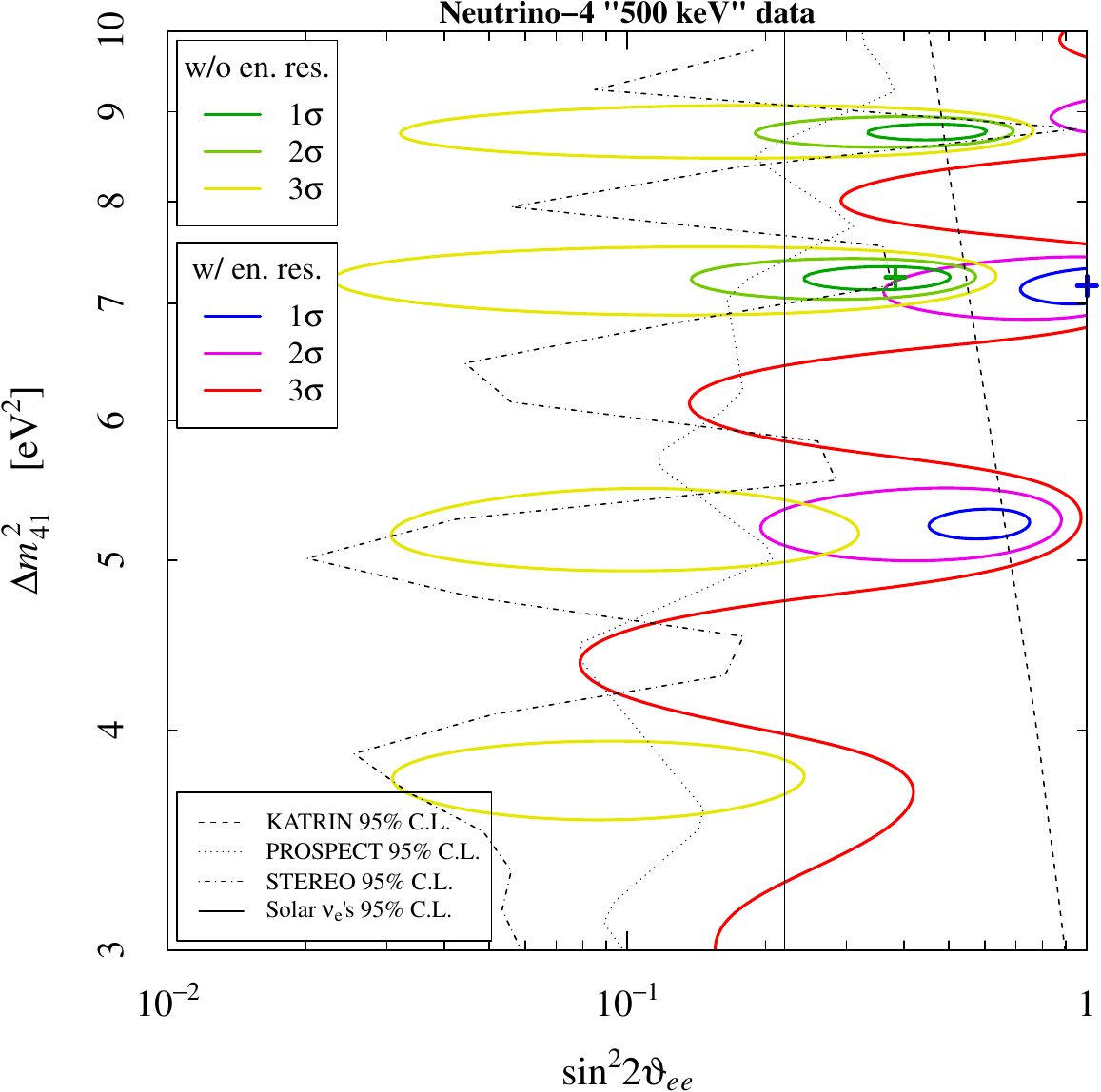}
\\
\end{tabular}
}
&
\subfigure[]{\label{fig:avg-Wilks}
\begin{tabular}{c}
\includegraphics*[width=0.49\linewidth]{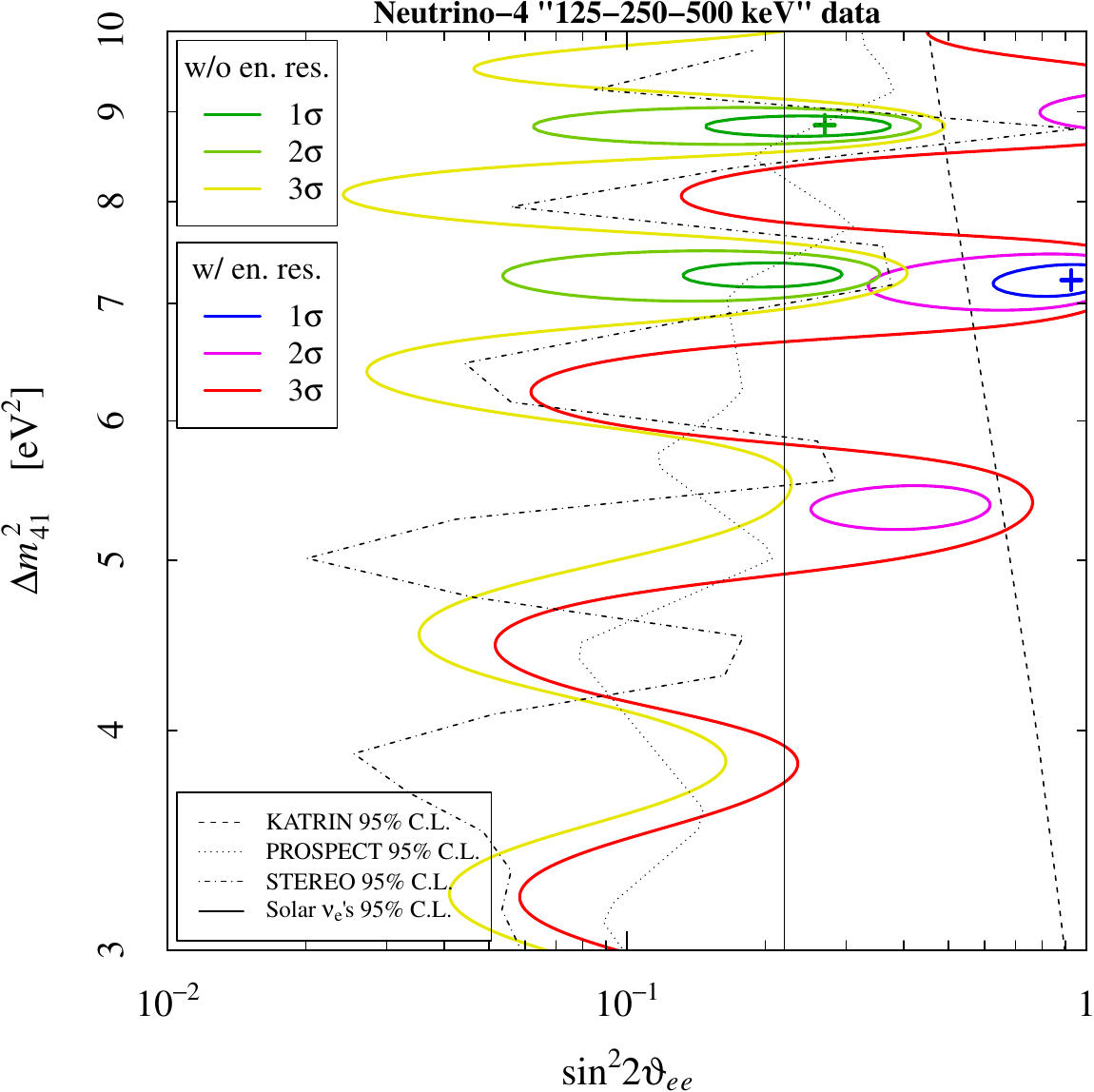}
\\
\end{tabular}
}
\end{tabular}
\caption{ \label{fig:Wilks}
Contours of the allowed regions in the
($\sin^2\!2\vartheta_{ee},\Delta{m}^2_{41}$)
plane obtained with the analyses without and with energy resolution of the Neutrino-4
\subref{fig:500-Wilks}
"500 keV"
and
\subref{fig:avg-Wilks}
"125-250-500 keV" data.
The allowed regions are calculated with the standard $\Delta\chi^2$ method
based on Wilks' theorem
($ \Delta\chi^2
=
\chi^2 - \chi^2_{\text{min}}
=
2.3, \, 6.2, \, 11.8
$
for
$
1\sigma, \, 2\sigma, \, 3\sigma
$, respectively).
The figures show also the 95\% C.L. exclusion curves of
KATRIN~\cite{Aker:2020vrf},
PROSPECT~\cite{Andriamirado:2020erz},
STEREO~\cite{AlmazanMolina:2019qul}, and
solar $\nu_{e}$'s
that were calculated with the same statistical method.
}
\end{figure*}

However,
the magenta histogram in Figure~\ref{fig:500keV} shows that
the energy smearing due to the energy resolution of the detector
suppresses strongly the oscillations at all values of $L/E$ and the data
cannot be fitted with the relatively small best-fit value of the effective mixing angle
claimed by the Neutrino-4 collaboration
($ \sin^2\!2\vartheta_{ee} = 0.38 $).
In order to fit the data taking into account
the strong suppression of the oscillations at all values of $L/E$
due to the energy resolution of the detector,
we need a value of the effective mixing angle that is close to our
maximal mixing best-fit value
($\sin^2\!2\vartheta_{ee}=1$),
as shown by the green histogram
in Figure~\ref{fig:500keV}.

The first two columns of Table~\ref{tab:fits}
give the results of our fits of the "500 keV" data
without and with the detector energy resolution.
Figure~\ref{fig:500-Wilks} shows a comparison of the corresponding allowed regions in the
($\sin^2\!2\vartheta_{ee},\Delta{m}^2_{41}$) plane.
We calculated these allowed regions with the standard $\Delta\chi^2$ method
based on Wilks' theorem
($ \Delta\chi^2
=
\chi^2 - \chi^2_{\text{min}}
=
2.3, \, 6.2, \, 11.8
$
for
$
1\sigma, \, 2\sigma, \, 3\sigma
$, respectively),
assuming, as usual, that the least-squares function (\ref{chi2})
is proportional to the log-likelihood.

From Figure~\ref{fig:500-Wilks},
one can see that the allowed regions and best-fit values of the oscillation parameters
that we obtained without taking into account
the detector energy resolution reproduce approximately
the results of the Neutrino-4 collaboration~\cite{Serebrov:2020kmd}\footnote{
The results that we obtained in Figure~\ref{fig:500-Wilks}
without taking into account the detector energy resolution
are also in agreement with the results obtained in Ref~\cite{Coloma:2020ajw}.
Let us, however, remark that the purpose of Ref~\cite{Coloma:2020ajw}
was not of checking or improving the fit of the Neutrino-4 data,
but of discussing the impact of the violation of
Wilks’ theorem on the
statistical significance of the Neutrino-4
oscillation signal.
It is reasonable to neglect the detector energy resolution
for that purpose.
The results of our investigation of the effects of the violation of
Wilks’ theorem presented in Section~\ref{sec:MC}
are in agreement with those in Ref.~\cite{Coloma:2020ajw}.
}.
Unfortunately,
these results are not realistic,
because the smearing due to the energy resolution of the detector
has a strong effect,
that moves the best fit to maximal mixing and enlarges
dramatically the allowed regions.
From Figure~\ref{fig:500-Wilks}
one can see that the $3\sigma$ allowed regions
with energy resolution are not bounded for small values of
$\sin^2\!2\vartheta_{ee}$,
in contrast to those obtained without energy resolution.
Therefore,
as one can see from Table~\ref{tab:fits},
taking into account the detector energy resolution
lowers the significance of the indication in favor of neutrino oscillations
obtained from the "500 keV" data from
$3.2\sigma$
to
$2.7\sigma$.

These $\sigma$ values have been obtained from the $p$-values
given in Table~\ref{tab:fits},
that have been calculated considering
for the difference of the value of $\chi^2$ without neutrino oscillations and
the value of $\chi^2_{\text{min}}$ in the case of neutrino oscillations
assuming a $\chi^2$ distribution with two degrees of freedom
(corresponding to the two fitted oscillation parameters
$\sin^22\vartheta_{ee}$ and $\Delta{m}^2_{41}$),
according to Wilks' theorem~\cite{Wilks:1938dza}.
A more reliable Monte Carlo estimation of the $p$-values and corresponding $\sigma$-values
is discussed in Section~\ref{sec:MC}.

Figure~\ref{fig:500-Wilks}
shows also the 95\% C.L. exclusion curves of KATRIN~\cite{Aker:2020vrf}
(see also Ref.~\cite{Giunti:2019fcj}),
PROSPECT~\cite{Andriamirado:2020erz},
STEREO~\cite{AlmazanMolina:2019qul}, and
solar $\nu_{e}$'s
that are calculated with the same method of the Neutrino-4 allowed regions.
One can see that all these exclusion curves
are in tension with the Neutrino-4 allowed regions.
In particular,
both the PROSPECT and STEREO exclusion curves
disfavor the Neutrino-4 $2\sigma$
regions obtained taking into account the energy resolution of the detector.
The KATRIN exclusion region disfavors the Neutrino-4 $1\sigma$ region
surrounding the best-fit point at maximal mixing.

We calculated a new solar neutrino bound on
$\sin^22\vartheta_{ee}$~\cite{Giunti:2009xz,Palazzo:2011rj,Palazzo:2012yf,Giunti:2012tn,Palazzo:2013me,Gariazzo:2017fdh}
considering only solar neutrino data,
without the KamLAND constraint,
that depends on the absolute reactor neutrino fluxes.
The degeneracy of the effects of $\vartheta_{ee}=\vartheta_{14}$
and $\vartheta_{13}$ is resolved by using the model-independent constraint
on $\vartheta_{13}$ obtained independently form the absolute reactor neutrino fluxes
by the
Daya Bay~\cite{Adey:2018zwh},
RENO~\cite{Bak:2018ydk},
and
Double Chooz~\cite{DoubleChooz:2019qbj}
reactor neutrino experiments
through the comparison of the data measured at near and far detectors.
Therefore,
the limit that we obtained
($\sin^22\vartheta_{ee} < 0.22$ at 95\% C.L.)
is model-independent and robust.
One can see from Figure~\ref{fig:500-Wilks}
that the solar neutrino bound disfavors
the large-mixing Neutrino-4 allowed regions.
Indeed,
such large values of %$\vartheta_{ee}=\vartheta_{14}$
3+1 active-sterile neutrino mixing would not be a minimal perturbation of
the standard three-neutrino mixing and
would spoil the standard three-neutrino mixing global fit of
solar, atmospheric and long-baseline (accelerator and reactor) neutrino oscillation
data~\cite{Capozzi:2017ipn,Esteban:2020cvm,deSalas:2020pgw}.

As already mentioned in Section~\ref{sec:method},
the Neutrino-4 collaboration considered also the values of
$R_{j}^{\text{exp}}$
averaged over 125, 250 and 500 keV energy bin widths,
arguing that the average of different samplings of the data
suppresses the fluctuations.
They presented the results obtained in this way as the most reliable results
of the experiment.
As we already remarked in Section~\ref{sec:method},
we disagree with this method because the procedure is rather ad-hoc
and it is not appropriate to consider
energy bin widths that are much smaller than the energy resolution of the detector.
However,
we are obliged to consider also this procedure in order to check the corresponding
results of the Neutrino-4 collaboration.

\begin{figure*}[!t]
\centering
\includegraphics*[width=\linewidth]{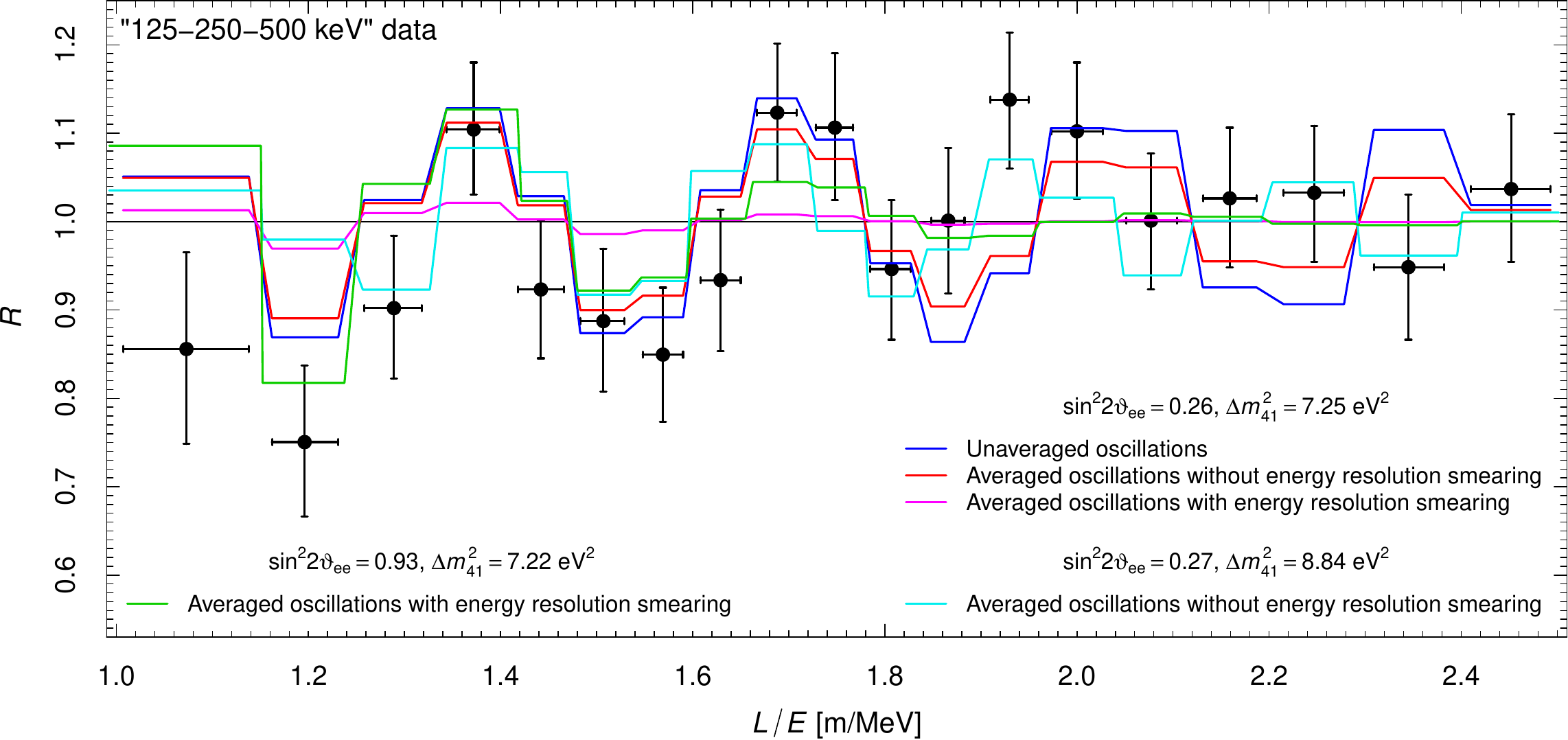}
\caption{ \label{fig:125-250-500keV}
Values of the
"125-250-500 keV" experimental ratios $R_{j}^{\text{exp}}$
(black points with error bars).
The blue, red, and magenta histograms
show, respectively,
the values of the theoretical ratios
$R_{j}^{\text{the}}$
obtained
for
($ \sin^2\!2\vartheta_{ee} = 0.26 $, $ \Delta{m}^2_{41} = 7.25 \, \text{eV}^2 $)
without any averaging of the oscillating terms,
with averaging over the energy and distance intervals without energy resolution,
and
with averaging over the energy and distance intervals with energy resolution.
The cyan and green histogram
correspond, respectively, to the best-fit values
($ \sin^2\!2\vartheta_{ee} = 0.93 $, $ \Delta{m}^2_{41} = 7.22 \, \text{eV}^2 $)
and
($ \sin^2\!2\vartheta_{ee} = 0.27 $, $ \Delta{m}^2_{41} = 8.84 \, \text{eV}^2 $)
obtained
with averaging over the energy and distance intervals without and with energy resolution
(see Table~\ref{tab:fits}).
}
\end{figure*}

As one can see from the fourth column in Table~\ref{tab:fits}
and from Figure~\ref{fig:avg-Wilks},
also using the "125-250-500 keV" data we obtained a large best-fit value of the
effective mixing angle,
that is smaller than the maximal best-fit value
obtained from the "500 keV" data,
but still very large,
with maximal mixing allowed within $1\sigma$.
This is what should be expected
if one considers energy bin widths that are smaller than the energy resolution of the detector:
there is slightly less averaging of the oscillations due to the
smaller energy bin widths,
but the dominant averaging effect due to the energy resolution of the detector
still requires the effective mixing angle to be very large
in order to fit the data.

On the other hand,
from the analysis of the "125-250-500 keV" data
the Neutrino-4 collaboration obtained the best-fit values
$ \sin^2\!2\vartheta_{ee} = 0.26 $
and
$ \Delta{m}^2_{41} = 7.25 \, \text{eV}^2 $
and the allowed regions in the
($\sin^2\!2\vartheta_{ee},\Delta{m}^2_{41}$)
plane were shown in Figure~56 of Ref.~\cite{Serebrov:2020kmd}.
However, according to our analysis this relatively small value of the effective mixing angle
cannot fit the data if the energy resolution of the detector is taken into account,
as illustrated by the magenta histogram in Figure~\ref{fig:125-250-500keV}.
Here we have a situation similar to what we have found for the "500 keV" data:
the "125-250-500 keV" data can be fitted with $ \sin^2\!2\vartheta_{ee} \simeq 0.26 $
only by neglecting the energy resolution of the detector.
Indeed,
as shown in Table~\ref{tab:fits} and Figure~\ref{fig:avg-Wilks},
neglecting the energy resolution of the detector
we obtained the best fit at
$ \sin^2\!2\vartheta_{ee} = 0.27 $
and
$ \Delta{m}^2_{41} = 8.8 \, \text{eV}^2 $,
and there is an allowed region at $1\sigma$ at
$ \sin^2\!2\vartheta_{ee} = 0.2 $
and
$ \Delta{m}^2_{41} \simeq 7.2 \, \text{eV}^2 $,
corresponding to the best fit of the Neutrino-4 collaboration.

Table~\ref{tab:fits} and Figure~\ref{fig:avg-Wilks}
show also that the indication in favor of neutrino oscillations
obtained with the "125-250-500 keV" data
is less significant than that obtained with the "500 keV" data
if the energy resolution of the detector is neglected
($2.7\sigma$
instead of
$3.2\sigma$),
but it is almost the same
when the energy resolution of the detector is taken into account
($2.8\sigma$
and
$2.7\sigma$).

The similarity of the statistical significances in favor of neutrino oscillations
that we obtained from the analyses of the "125-250-500 keV" data
without and with the energy resolution of the detector
($2.7\sigma$
and
$2.8\sigma$,
respectively)
is due to the similarity of the quality of the best fit in the two cases,
that correspond to almost equal values of $\chi^2_{\text{min}}$,
as shown in Table~\ref{tab:fits}.
From Figure~\ref{fig:125-250-500keV}
one can see that the fit of the data in the two cases is different,
but in both cases it is not very good for some data points.

Figure~\ref{fig:avg-Wilks}
shows that also considering the "125-250-500 keV" data
the PROSPECT, STEREO, and solar exclusion curves
disfavor the Neutrino-4 $2\sigma$
regions obtained taking into account the energy resolution of the detector,
and
the KATRIN exclusion region disfavors the Neutrino-4 $1\sigma$ region
surrounding the best-fit point at maximal mixing.

In conclusion of this Section, it is useful to summarize our findings:

\begin{enumerate}

\item
Both the "500 keV" and the "125-250-500 keV" data of the Neutrino-4 experiment
can be fitted well with the values of the effective mixing angle
claimed by the Neutrino-4 collaboration
($ \sin^2\!2\vartheta_{ee} = 0.38 $ for the "500 keV"
and
$ \sin^2\!2\vartheta_{ee} = 0.26 $ for the "125-250-500 keV")
by neglecting the energy resolution of the detector.

\item
If the energy resolution of the detector is taken into account,
the best fit value of the effective mixing angle is maximal
($ \sin^2\!2\vartheta_{ee} = 1 $)
for the "500 keV" data
and close to maximal
($ \sin^2\!2\vartheta_{ee} = 0.93 $)
for the "125-250-500 keV" data.

\item
If the energy resolution of the detector is taken into account,
the indication
in favor of short-baseline oscillations
of the Neutrino-4 data
is about
$2.7-2.8\sigma$
if one assumes the approximate validity of Wilks' theorem.

\item
If the energy resolution of the detector is taken into account,
there is a strong tension between the Neutrino-4 allowed regions
in the
($\sin^2\!2\vartheta_{ee},\Delta{m}^2_{41}$)
plane
and
the 95\% C.L. exclusion curves of
KATRIN~\cite{Aker:2020vrf},
PROSPECT~\cite{Andriamirado:2020erz},
STEREO~\cite{AlmazanMolina:2019qul}, and
solar $\nu_{e}$'s.

\end{enumerate}

In Section~\ref{sec:MC} we present a more reliable Monte Carlo estimation of the
statistical significance of the indication
in favor of short-baseline oscillations
of the Neutrino-4 data
and the corresponding estimation of the
allowed regions in the
($\sin^2\!2\vartheta_{ee},\Delta{m}^2_{41}$)
plane.

\begin{figure*}[!t]
\centering
\setlength{\tabcolsep}{0pt}
\begin{tabular}{cc}
\subfigure[]{\label{fig:500-nores-MC}
\begin{tabular}{c}
\includegraphics*[width=0.49\linewidth]{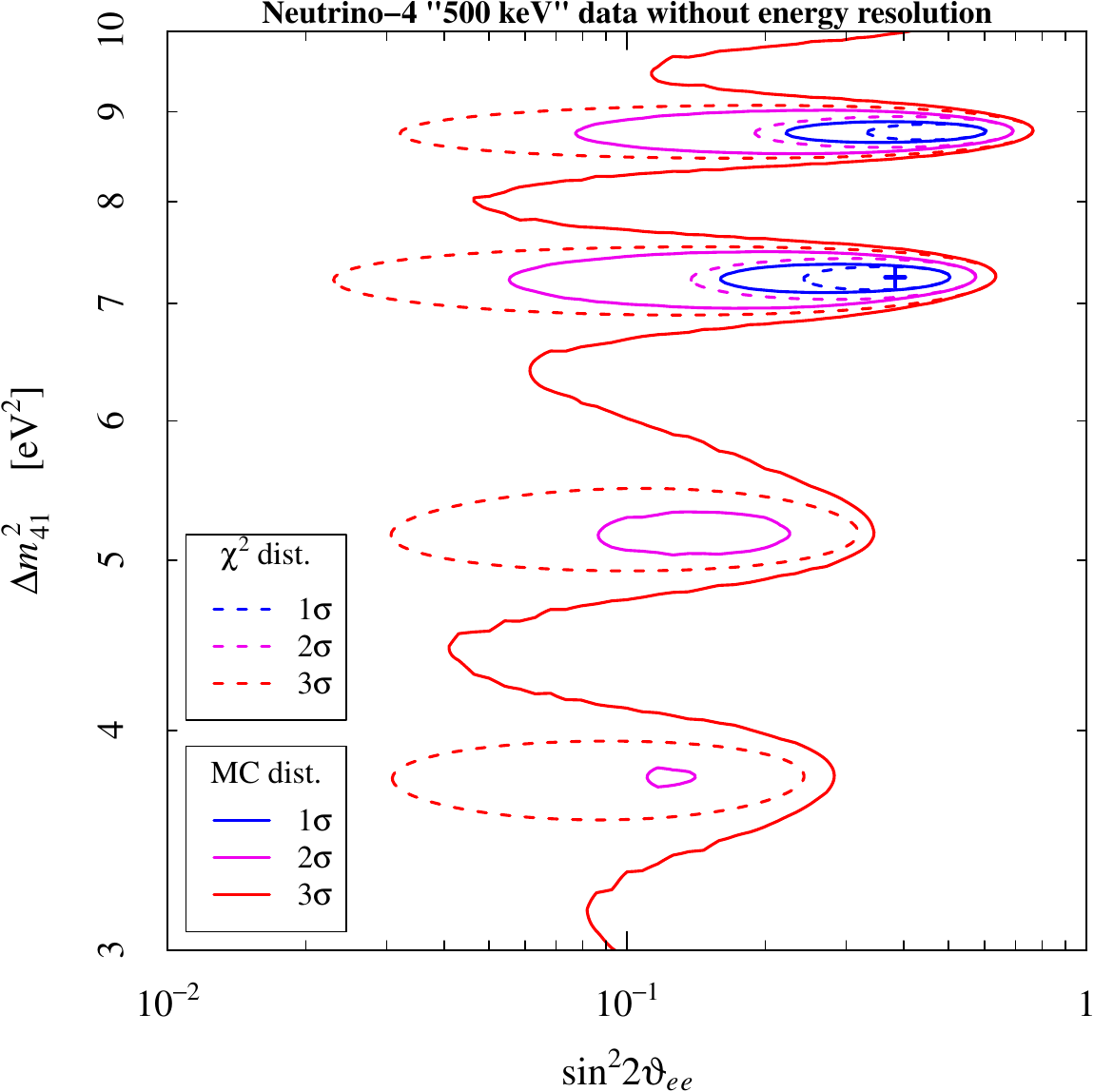}
\\
\end{tabular}
}
&
\subfigure[]{\label{fig:500-resol-MC}
\begin{tabular}{c}
\includegraphics*[width=0.49\linewidth]{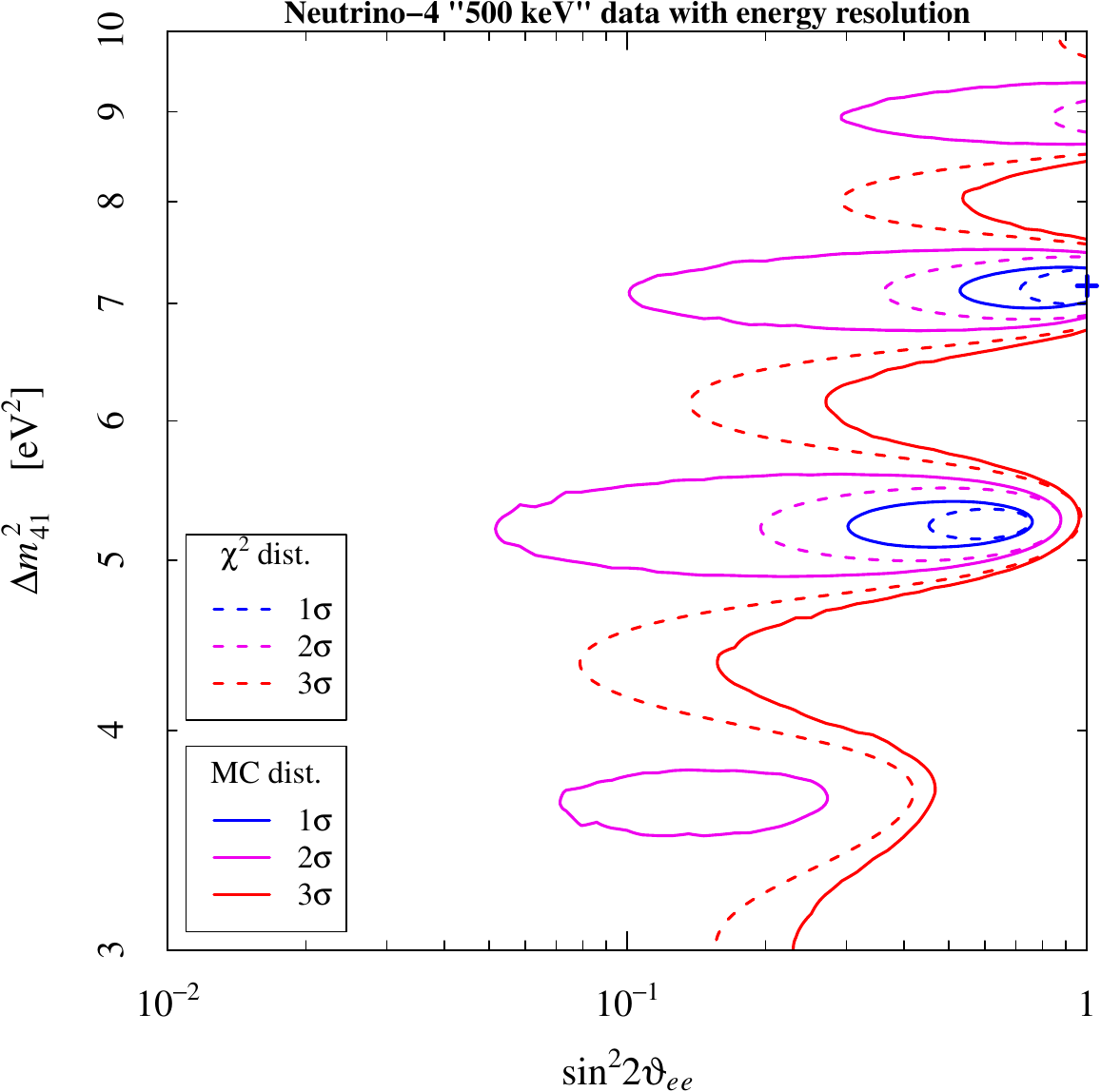}
\\
\end{tabular}
}
\\
\subfigure[]{\label{fig:avg-nores-MC}
\begin{tabular}{c}
\includegraphics*[width=0.49\linewidth]{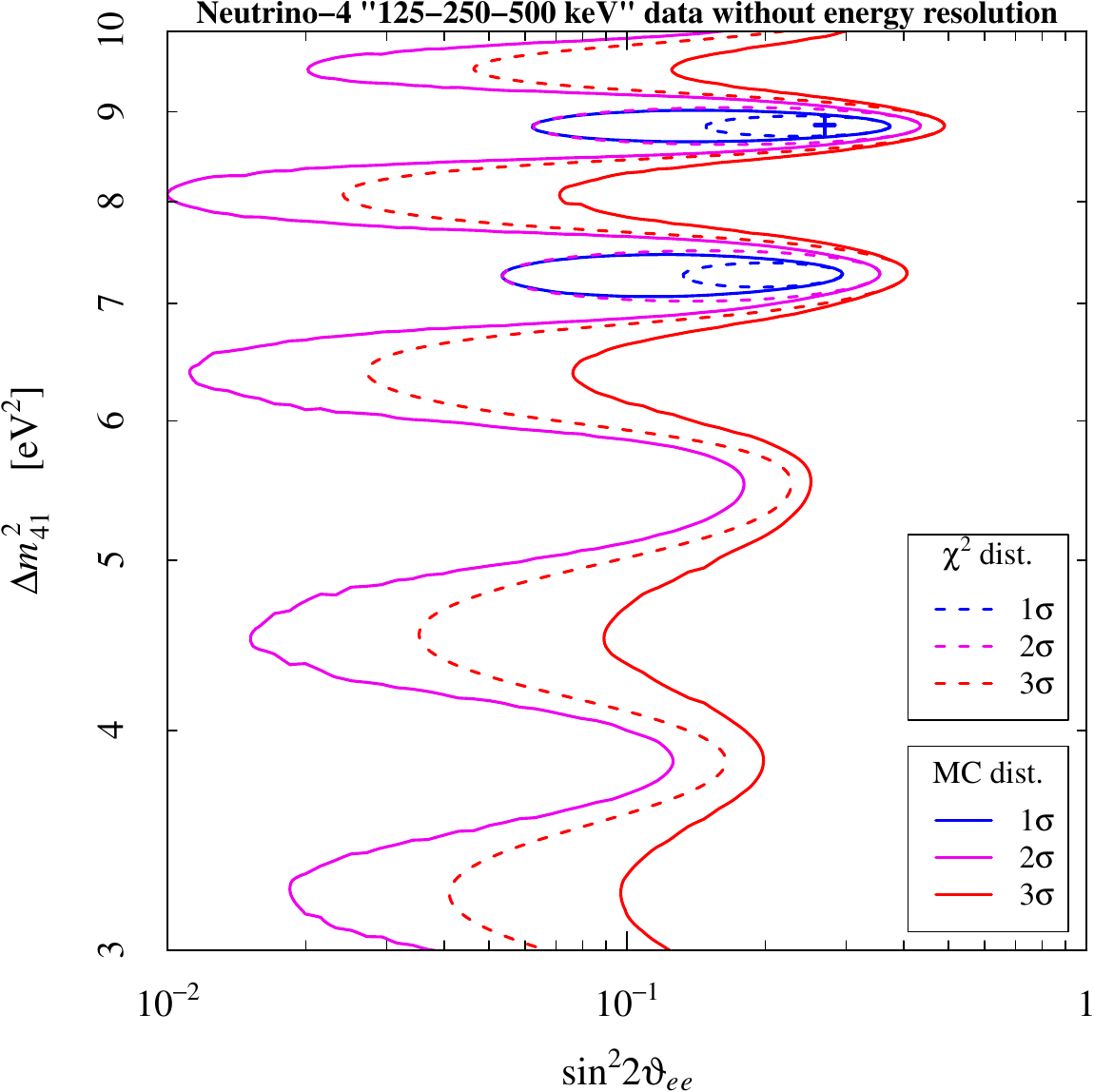}
\\
\end{tabular}
}
&
\subfigure[]{\label{fig:avg-resol-MC}
\begin{tabular}{c}
\includegraphics*[width=0.49\linewidth]{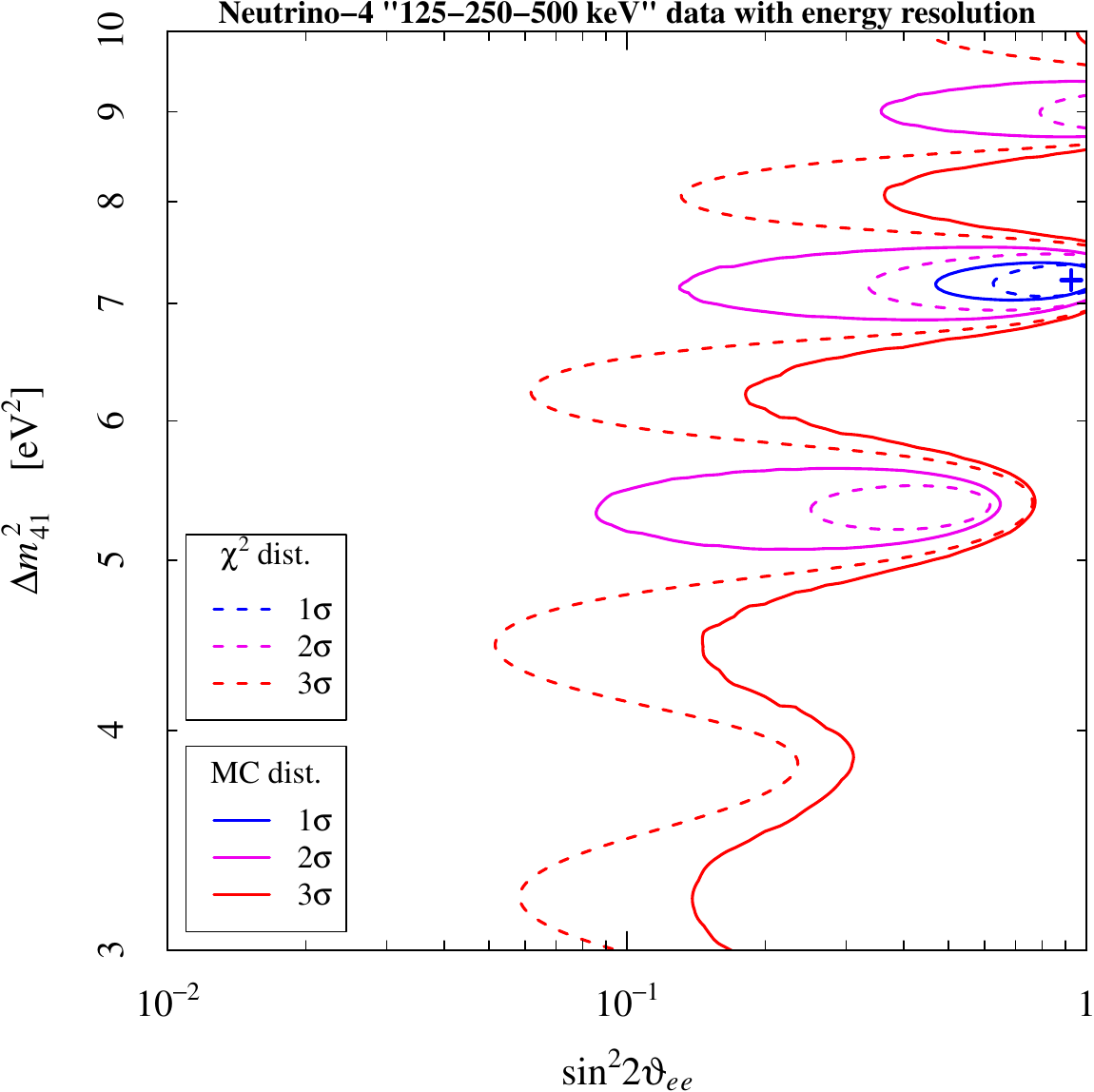}
\\
\end{tabular}
}
\end{tabular}
\caption{ \label{fig:MC}
Contours of the allowed regions in the
($\sin^2\!2\vartheta_{ee},\Delta{m}^2_{41}$)
plane obtained with the analyses of the Neutrino-4
"500 keV" data
\subref{fig:500-nores-MC}
without and
\subref{fig:500-resol-MC}
with energy resolution,
and those obtained from the "125-250-500 keV"
\subref{fig:avg-nores-MC}
without and
\subref{fig:avg-resol-MC}
with energy resolution.
The solid lines represent the Monte Carlo evaluated contours,
that are confronted with the dashed lines
representing the contours in Figure~\ref{fig:Wilks} of the allowed regions
calculated assuming a $\chi^2$ distribution for
$
\Delta\chi^2
=
\chi^2
-
\chi^2_{\text{min}}
$.
}
\end{figure*}

\section{Monte Carlo confidence intervals}
\label{sec:MC}

It is well known that the conditions for the validity of Wilks' theorem are not always satisfied in neutrino oscillation experiments~\cite{Feldman:1997qc,Lyons:2014kta,Agostini:2019jup,Algeri:2019arh,Giunti:2020uhv,Coloma:2020ajw}.
In this case,
a Monte Carlo statistical analysis gives
a more reliable estimation of the confidence intervals for the oscillation parameters.
In this Section we present the results of Monte Carlo statistical analyses
of the
"500 keV" and the "125-250-500 keV" data of the Neutrino-4 experiment
without and with the effect of the energy resolution of the detector.

For each point on a grid in the
($\sin^2\!2\vartheta_{ee},\Delta{m}^2_{41}$)
plane
we generated a large number of random data sets (of the order of $10^5$)
with the uncertainties of the Neutrino-4 data set.
For each random data set we calculated the value of $\chi^2$
corresponding to the generating values of
$\sin^2\!2\vartheta_{ee}$ and $\Delta{m}^2_{41}$
and
we found the minimum value of $\chi^2$
in the ($\sin^2\!2\vartheta_{ee},\Delta{m}^2_{41}$)
plane.
Denoting these two quantities by
$\chi^2_{\text{MC}}(\sin^2\!2\vartheta_{ee},\Delta{m}^2_{41})$
and
$\chi^2_{\text{MC,min}}(\sin^2\!2\vartheta_{ee},\Delta{m}^2_{41})$,
we obtained the distribution of
$
\Delta\chi^2_{\text{MC}}(\sin^2\!2\vartheta_{ee},\Delta{m}^2_{41})
=
\chi^2_{\text{MC}}(\sin^2\!2\vartheta_{ee},\Delta{m}^2_{41})
-
\chi^2_{\text{MC,min}}(\sin^2\!2\vartheta_{ee},\Delta{m}^2_{41})
$,
that allows us to determine
if the value of
$
\Delta\chi^2(\sin^2\!2\vartheta_{ee},\Delta{m}^2_{41})
=
\chi^2(\sin^2\!2\vartheta_{ee},\Delta{m}^2_{41})
-
\chi^2_{\text{min}}(\sin^2\!2\vartheta_{ee},\Delta{m}^2_{41})
$
obtained with the analysis of the actual Neutrino-4 data
is included or not in a region with a fixed confidence level.

Figure~\ref{fig:MC}
shows the comparison of the
contours of the Monte Carlo allowed regions
with the Wilks $\Delta\chi^2$ contours
that were obtained in Figure~\ref{fig:Wilks} assuming a $\chi^2$ distribution for
$
\Delta\chi^2
=
\chi^2
-
\chi^2_{\text{min}}
$.
One can see that in all the four analyses
the Monte Carlo allowed regions are much larger than the Wilks $\Delta\chi^2$ allowed regions.
In particular,
for a fixed value of the confidence level
smaller values of the effective mixing are allowed by the Monte Carlo calculation
and
the absence of neutrino oscillation is compatible with the data at $3\sigma$
in all the four analyses.
This result is reflected in the decrease of the statistical significance
of the Neutrino-4 indication in favor of neutrino oscillations
shown in Table~\ref{tab:fits}.

Figure~\ref{fig:500-nores-MC} is in approximate agreement with Fig.~6 of Ref.~\cite{Coloma:2020ajw}.
The corresponding decrease of the statistical significance
of the Neutrino-4 oscillation signal from
$3.2\sigma$ assuming the validity of Wilks’ theorem to
the Monte Carlo $2.5\sigma$
(see Table~\ref{tab:fits})
is in good agreement with the decrease from
$3.2\sigma$ to $2.6\sigma$
found in Ref.~\cite{Coloma:2020ajw}.

Considering the more accurate analyses that take into account the energy resolution of the detector,
one can see that the Monte Carlo calculation lowers the
statistical significance
of the indication in favor of neutrino oscillations of the "500 keV" data
from
$2.7\sigma$
to
$2.2\sigma$
and that of the "125-250-500 keV" data
from
$2.8\sigma$
to
$2.2\sigma$.

\begin{figure*}[!t]
\centering
\setlength{\tabcolsep}{0pt}
\begin{tabular}{cc}
\subfigure[]{\label{fig:500-nores-sym}
\begin{tabular}{c}
\includegraphics*[width=0.49\linewidth]{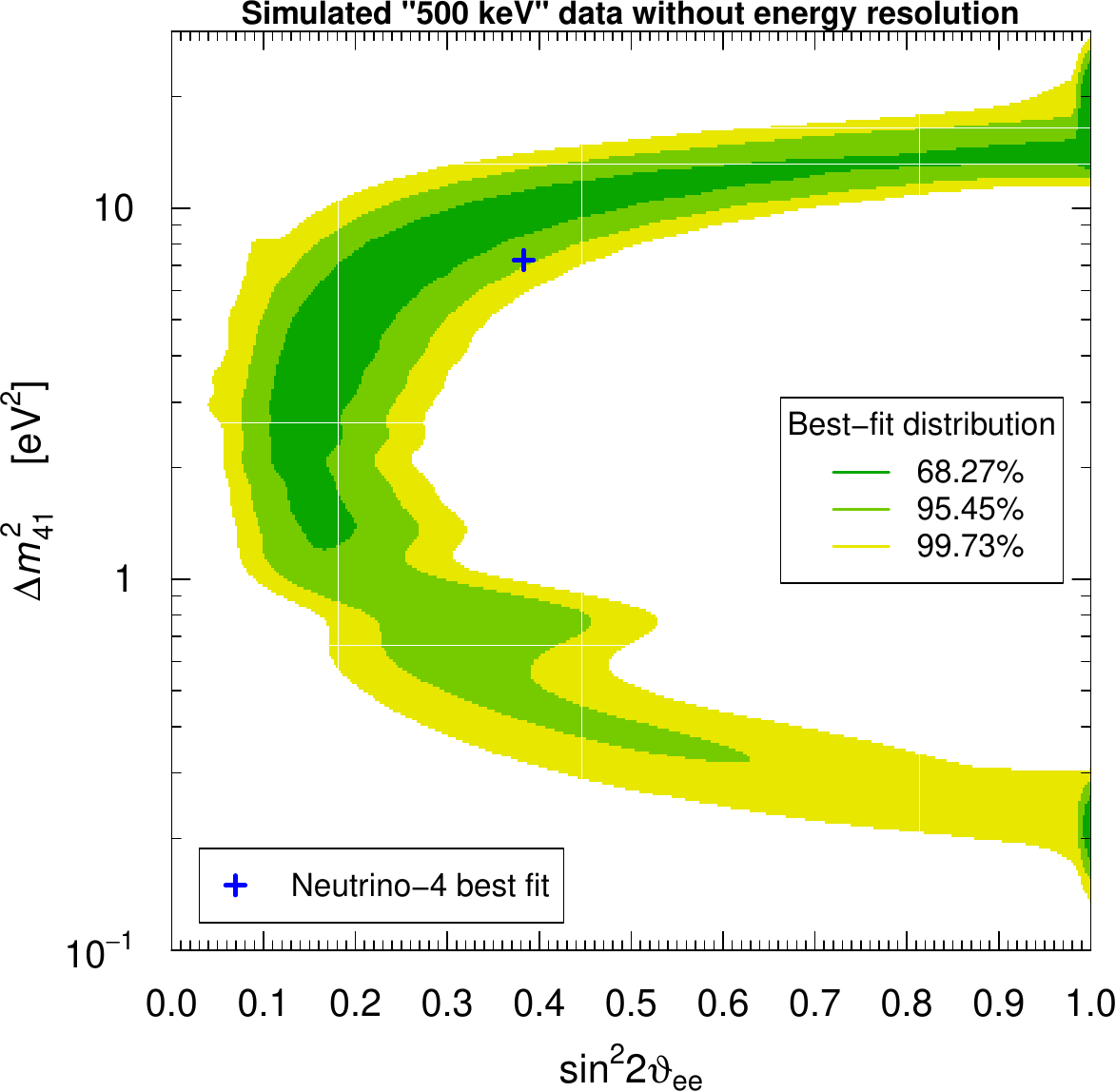}
\\
\end{tabular}
}
&
\subfigure[]{\label{fig:500-resol-sym}
\begin{tabular}{c}
\includegraphics*[width=0.49\linewidth]{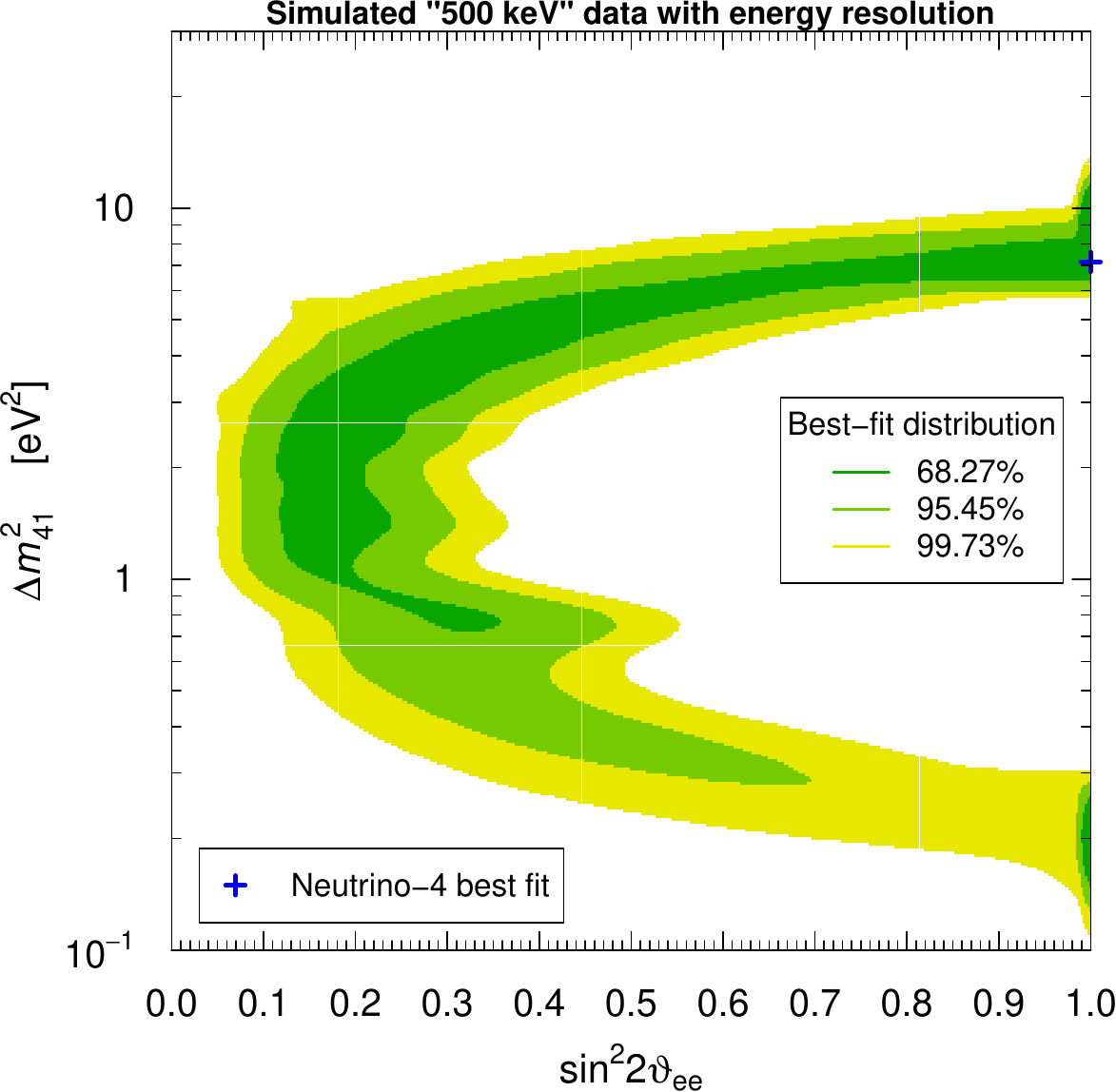}
\\
\end{tabular}
}
\\
\subfigure[]{\label{fig:avg-nores-sym}
\begin{tabular}{c}
\includegraphics*[width=0.49\linewidth]{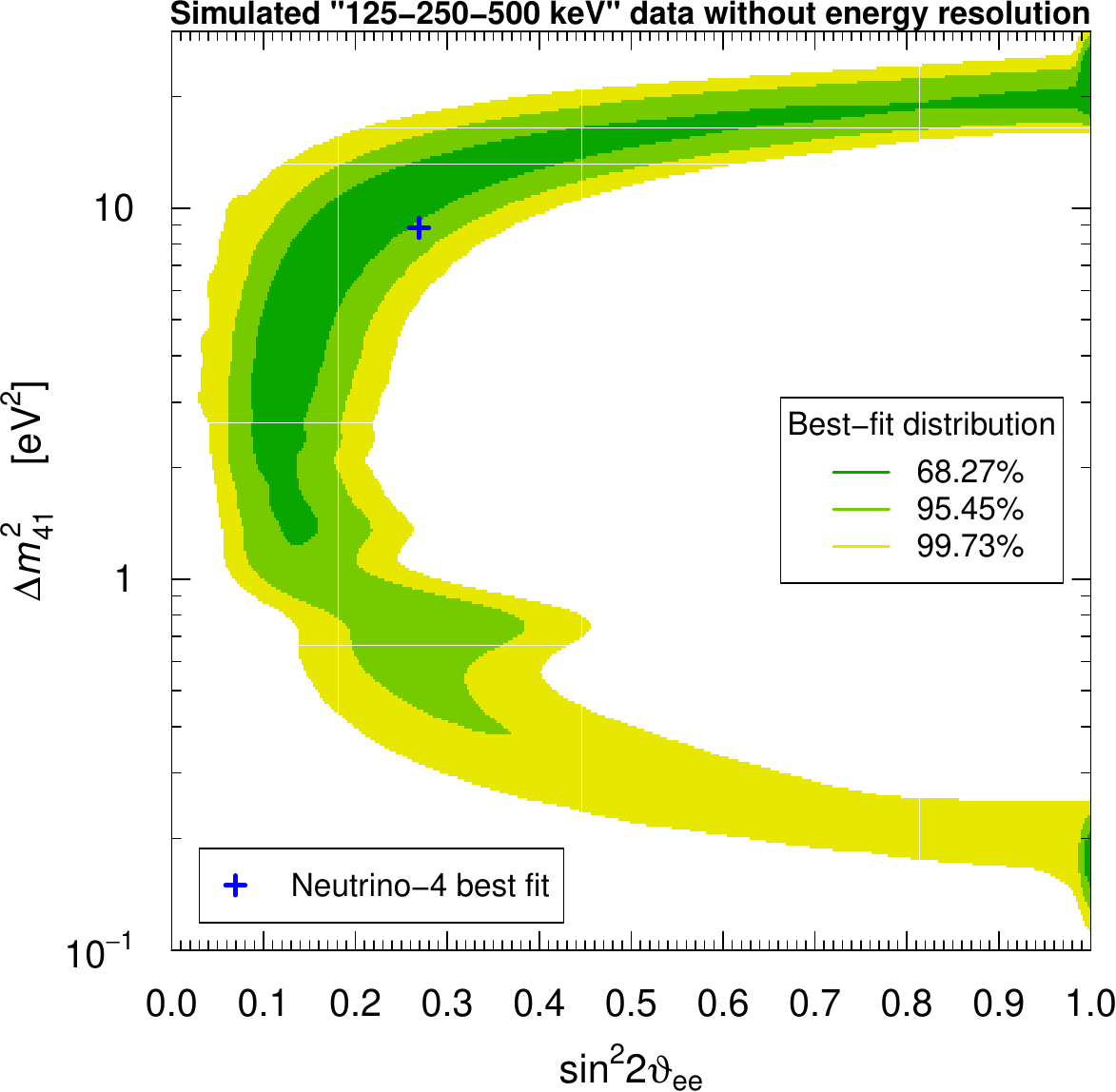}
\\
\end{tabular}
}
&
\subfigure[]{\label{fig:avg-resol-sym}
\begin{tabular}{c}
\includegraphics*[width=0.49\linewidth]{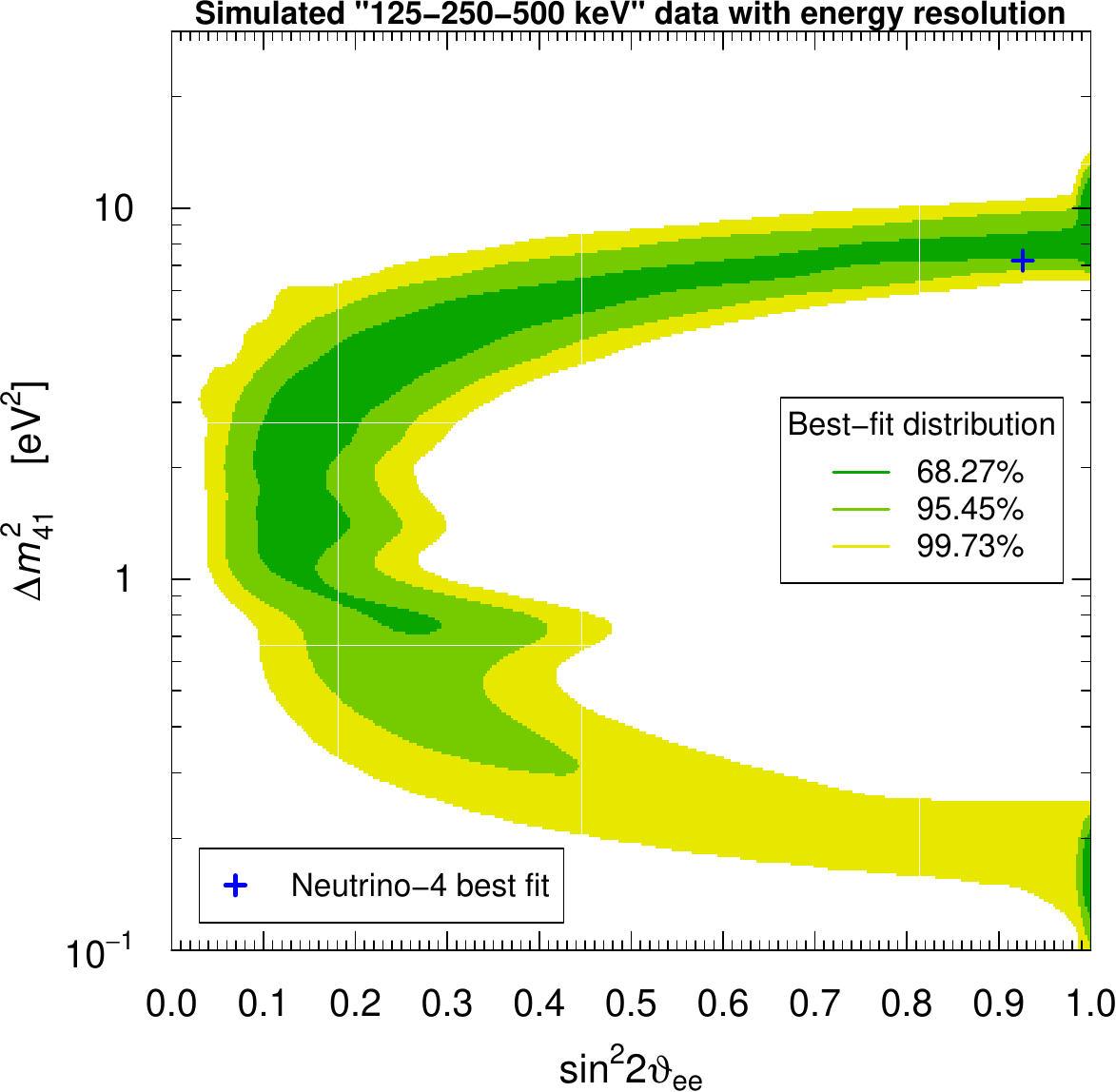}
\\
\end{tabular}
}
\end{tabular}
\caption{ \label{fig:sym}
Distribution of the best-fit points
in the absence of neutrino oscillations
obtained with simulated data corresponding to the Neutrino-4
or data analyzed
\subref{fig:500-nores-sym}
without and
\subref{fig:500-resol-sym}
with energy resolution,
and to the Neutrino-4 "125-250-500 keV" analyzed
\subref{fig:avg-nores-sym}
without and
\subref{fig:avg-resol-sym}
with energy resolution.
The blue crosses indicate the actual best fit points obtained
from the Neutrino-4 data
(already reported in Tab.~\ref{tab:fits}
and Figures~\ref{fig:Wilks} and~\ref{fig:MC}).
}
\end{figure*}

\begin{table*}[t!]
\centering
\renewcommand{\arraystretch}{1.2}
\begin{tabular}{ccccc}
&
\multicolumn{2}{c}{"500 keV" data}
&
\multicolumn{2}{c}{"125-250-500 keV" data}
\\[-0.3cm]
&
\multicolumn{2}{c}{\rule{2.5cm}{0.5pt}}
&
\multicolumn{2}{c}{\rule{2.5cm}{0.5pt}}
\\
&
{\renewcommand{\arraystretch}{0.8}
\begin{tabular}{c}
without
\\
en. res.
\end{tabular}
}
&
{\renewcommand{\arraystretch}{0.8}
\begin{tabular}{c}
with
\\
en. res.
\end{tabular}
}
&
{\renewcommand{\arraystretch}{0.8}
\begin{tabular}{c}
without
\\
en. res.
\end{tabular}
}
&
{\renewcommand{\arraystretch}{0.8}
\begin{tabular}{c}
with
\\
en. res.
\end{tabular}
}
\\
\hline
$P(\sin^2\!2\vartheta_{ee}<0.1)$
&
$0.009$
&
$0.008$
&
$0.044$
&
$0.031$
\\
\hline
$P(0.1<\sin^2\!2\vartheta_{ee}<0.5)$
&
$0.680$
&
$0.625$
&
$0.708$
&
$0.645$
\\
\hline
$P(0.5<\sin^2\!2\vartheta_{ee}<0.9)$
&
$0.152$
&
$0.184$
&
$0.142$
&
$0.163$
\\
\hline
$P(\sin^2\!2\vartheta_{ee}>0.9)$
&
$0.159$
&
$0.183$
&
$0.105$
&
$0.161$
\end{tabular}
\caption{ \label{tab:sym}
Probabilities to obtain the best-fit in four ranges of $\sin^2\!2\vartheta_{ee}$
in the absence of neutrino oscillations
with the four methods of data analysis discussed in the paper.
}
\end{table*}

Therefore,
although
our analyses of the Neutrino-4 data show that
taking into account the energy resolution of the detector
leads to a best-fit value of the mixing close to maximal,
that is in contradiction with the bounds of other experiments,
the allowed range of the mixing is large and includes
the absence of mixing, and hence of oscillations,
at
$2.2\sigma$,
independently on the choice of the
"500 keV" or "125-250-500 keV" data set,
with a reliable Monte Carlo calculation.
Hence,
it is unclear if the Neutrino-4 anomaly is due to neutrino oscillations
or a statistical fluctuation of the data.

A further indication can be obtained by finding the distribution
of the best-fit points assuming the absence of oscillations,
as done in Figure~1 of Ref.~\cite{Almazan:2020drb} for the STEREO experiment
and in Figure~2 of Ref.~\cite{Coloma:2020ajw}
for a toy reactor neutrino oscillation experiment.
Figure~\ref{fig:sym}
shows the distributions of best-fit points that we obtained
for the four analysis methods of the Neutrino-4 data
that we considered in this paper.
One can see that
the actual best fit points obtained from the Neutrino-4 data,
shown by the blue crosses,
lie in regions with high probability.
Therefore,
they are not unlikely in the absence of oscillations,
in spite of their large mixing.

From Figure~\ref{fig:sym} one can also see that,
although the data are generated assuming the absence of mixing,
the probability of finding zero-mixing best fit is negligible.
Instead, there is a large probability to find large values of the mixing,
in the range
$ 0.1 \lesssim \sin^2\!2\vartheta_{ee} \leq 1 $.
Table~\ref{tab:sym} shows the values of the probabilities
in four intervals of $\sin^2\!2\vartheta_{ee}$.
One can see that there is a maximum of about 60-70\% of probability
to find the best fit for
$ 0.1 < \sin^2\!2\vartheta_{ee} < 0.5 $,
but the probability to find a best fit with
$ \sin^2\!2\vartheta_{ee} > 0.9 $
is far from negligible,
being about 16-18\% in the analyses taking into account the energy resolution of the detector.
Therefore,
it is clear that the large mixing of the best-fit points obtained from the analysis of the Neutrino-4 data is
compatible with the absence of oscillations.

\section{Conclusions}
\label{sec:Conclusions}

In this paper we have presented the results of
detailed analyses of the Neutrino-4 data
aimed at finding the significance of the
large-mixing
short-baseline neutrino oscillation signal claimed by the Neutrino-4
collaboration
at more than $3\sigma$~\cite{Serebrov:2018vdw,Serebrov:2020rhy,Serebrov:2020kmd}.
We found that the results of the Neutrino-4
collaboration can be reproduced approximately only by neglecting the
effects of the energy resolution of the detector.
Including these effects,
we found that the best-fit point and the surrounding $1\sigma$ allowed region in the
($\sin^2\!2\vartheta_{ee},\Delta{m}^2_{41}$)
plane lie at even larger values of the mixing than that claimed by the Neutrino-4
collaboration.
However,
the $3\sigma$ allowed region calculated with the standard $\Delta\chi^2$ method based on Wilks' theorem
is much larger than that claimed by the Neutrino-4
collaboration and include the case of zero mixing,
i.e. the absence of oscillations.
The corresponding statistical significance of
short-baseline neutrino oscillations is about
$2.7\sigma$.

We have also shown that
there is a strong tension between the Neutrino-4 allowed regions
in the
($\sin^2\!2\vartheta_{ee},\Delta{m}^2_{41}$)
plane
and
the 95\% C.L. exclusion curves of
KATRIN~\cite{Aker:2020vrf},
PROSPECT~\cite{Andriamirado:2020erz},
STEREO~\cite{AlmazanMolina:2019qul}, and
solar $\nu_{e}$'s.

We have further studied the statistical significance of the Neutrino-4
indication in favor of short-baseline neutrino oscillations
with a more reliable Monte Carlo evaluation of the distribution of
$\Delta\chi^2=\chi^2-\chi^2_{\text{min}}$
in the ($\sin^2\!2\vartheta_{ee},\Delta{m}^2_{41}$)
plane.
We found that the allowed regions extend to lower values of the mixing,
as expected~\cite{Agostini:2019jup,Giunti:2020uhv,Coloma:2020ajw},
and the statistical significance of
short-baseline neutrino oscillations decreases to about
$2.2\sigma$.

We have also shown with a Monte Carlo simulation of a large set of Neutrino-4-like data
that it is not unlikely to
obtain a best-fit point that has a large mixing,
even maximal,
in the absence of oscillations.
Therefore,
we conclude that the claimed Neutrino-4 indication in favor of short-baseline neutrino oscillations with very large mixing is rather doubtful.

\begin{acknowledgments}
We would like to thank G. Ranucci for useful comments on the first version of the paper.
The work of C. Giunti and C.A. Ternes was supported by the research grant "The Dark Universe: A Synergic Multimessenger Approach" number 2017X7X85K under the program PRIN 2017 funded by the Ministero dell'Istruzione, Universit\`a e della Ricerca (MIUR).
The work of Y.F. Li and Y.Y. Zhang is supported by the National Natural Science Foundation of China under Grant No.~12075255 and No.~11835013,
and by Beijing Natural Science Foundation under Grant No.~1192019.
Y.F. Li is also grateful for the support by the CAS Center for Excellence in Particle Physics (CCEPP).
\end{acknowledgments}

%merlin.mbs apsrev4-1.bst 2010-07-25 4.21a (PWD, AO, DPC) hacked
%Control: key (0)
%Control: author (72) initials jnrlst
%Control: editor formatted (1) identically to author
%Control: production of article title (-1) disabled
%Control: page (0) single
%Control: year (1) truncated
%Control: production of eprint (0) enabled
%
%\bibliographystyle{apsrev4-1}
%\bibliography{main,wrk/carlo}

\end{document}